\documentclass[aps,prl,twocolumn,groupedaddress,showpacs]{revtex4-1}

\usepackage{graphicx,color}
\usepackage{epsfig}
\usepackage{dcolumn}
\usepackage{amsmath,amssymb,bm}
\usepackage{hyperref,url}
\usepackage[noabbrev]{cleveref}
\usepackage{CJK}
\usepackage{lineno}
\usepackage[english]{babel}
\usepackage[T1]{fontenc}
\usepackage{pifont}
\usepackage{amsfonts}
\usepackage{wasysym}
\usepackage{amsbsy}
\usepackage{psfrag}
\usepackage{color}
\usepackage{multirow}
\usepackage{cases}
\usepackage{stackengine}
\usepackage{float}
\usepackage{blkarray}
\usepackage{cancel}
\usepackage{mathrsfs}
\usepackage{empheq}
\usepackage{mathptmx} 

\begin{document}

\preprint{APS/123-QED}

\title{\textbf{Digital Sub-millimeter Bubble-Jets} 
}%

\author{B. J. Ruan}
\author{{Z. L. Wang}}%
 \email{wng_zh@i.shu.edu.cn}
\affiliation{Shanghai Institute of Applied Mathematics and Mechanics, Shanghai Key Laboratory of Mechanics in Energy Engineering, Shanghai University, Yanchang Road 149, Shanghai, 200072, P.R. China
}%

\date{\today}

\begin{abstract}

We create digital sub-millimeter bubble-jet emitting in gas-liquid co-flows at tapered chip zone for moderate $Re \sim [20, 120]$. Self-similarity features are revealed at the tapered area and giving birth to a local model. Local self-scaling characteristic quantities, $W_{\text{local}}$ and $L_{\text{cone}}$, are introduced to scale energies and progresses at onsets of bubble-jets, which gives highly universal phase diagram and also scaling law of jet velocity. The phase diagram draws critical bubble-jetting line at $We_d\sim Ca_c^{-5.7}$ and jet-dropping line at $We_d\sim Ca_c^{-4.2}$, as well as orthogonally overlaping Taylor bubble area from annular flow pattern. And the jetting velocity expresses as $u_{\text{jet}}\sim [\rho_c^2\sigma (Q_c + Q_d)^2]/(\mu_c^3 W_{\text{local}}^{1.6}H^{0.4})$, which clarifies the compound mechanisms for bubble-jet emitting by combined competing of interfacial tension, inertia, viscosity, and the local tapered geometries. These universalities confirm reciprocally similarities of the bubble-jet emitting processes on behaviors and flow structures at the local tapered zone.
 
\end{abstract}


\pacs{47.55.Dz 47.62.+q 47.55.Kf}              
\maketitle

Sub-millimeter bubble-jet is an uncommon phenomenon that shows great potential in fields such as drug delivery \cite{rodriguez2015generation}, biochemical reactions \cite{patel2021advances}, and cell puncture \cite{nakagawa2017pulsed}. Many studies trigger cavitation through external excitations like ultrasound \cite{lauterborn2023acoustic,yusof2016physical}, shock waves \cite{deng2024impact,huang2021physical}, or lasers \cite{chen2024laser,agrevz2024laser}, thereby inducing sub-millimeter bubble-jet. However, due to the complex applied conditions, the stability and controllability of the jet are relatively too low to show clear practical prospects.Researches on bubble-jet above millimeters are common, typically focus on bubble emergence \cite{brasz2018minimum,cheng2020numerical,berny2020role}, bubble coalescence \cite{andredaki2021accelerating}, and bubble stretching \cite{seon2012large,li2019jet}, or gravity-driven \cite{cheng2020numerical,berny2020role,seon2012large} and surface-wave focusing \cite{li2019jet} bubble-jets, etc. Unlikely, these studies are macroscopically related to natural fields like oceanic atomization, aerosol generation, atmospheric bubble dynamics, cloud - formation processes, and air-pollution control \cite{blanco2020sea,lorenceau2004air,liger2003capillary,cochran2017sea,chingin2018enrichment}, gravity often cannot be ignored or such phenomena occur in open areas with free boundaries. The sub-millimeter bubble-jets limit to the capillary scale ($l_c=\sqrt{\sigma/(\rho g)}\approx 2.7mm$), usually $Bo < 1$ ($Bo=\rho g L^2/\sigma$). The gravitation is relatively fail to surface effects, and capillary force arouses to restrain gas-liquid deformation. Therefore, the sub-millimeter bubble-jet usually requires externalities stimulating as mentioned above. Ultrasound \cite{nikolov2019air,tan2009interfacial} triggers bubble oscillation and cavitation through periodic pressure waves, leading to bubble expansion and rupture; Shock waves \cite{luo2019jet} drive the rapid compression and expansion through drastic pressure changes; Lasers \cite{brown2011time,patrascioiu2014laser} rapidly heat the bubbles through the photothermal effect, generating rapid internal pressure differences. While in field of digital microfluidics, since usually $Re < 1$, the bubble-jet phenomenon is rarely involved. Few research may be related, such as inward-rolling bubble-jets at relatively large Reynolds number ($Re\approx 400$) \cite{pico2024drop},  however, research on generating digital sub-millimeter bubble-jets has not reported yet.

\begin{figure}[!t]
  \centering
  \includegraphics[width=8.5cm,keepaspectratio]{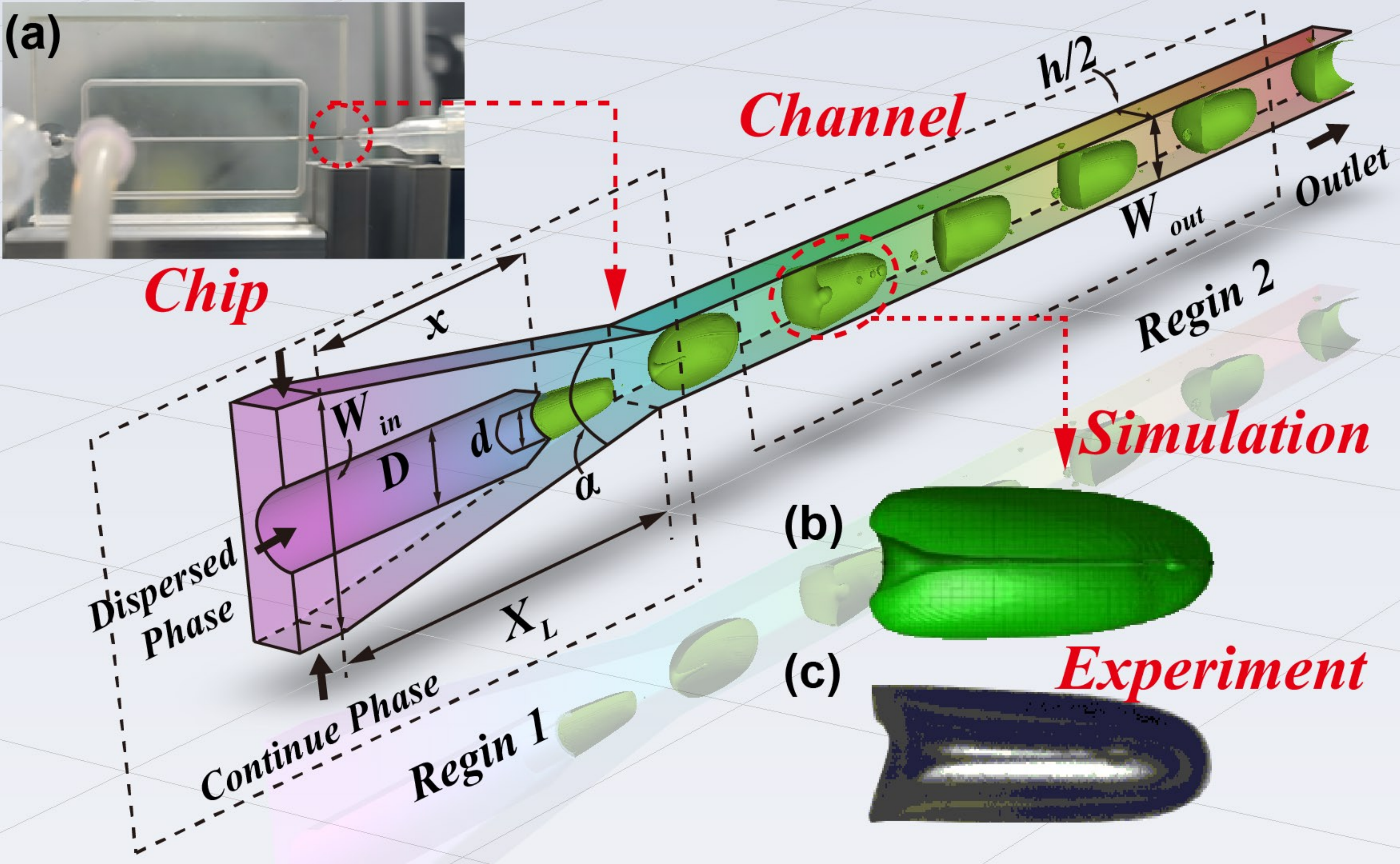}
  \caption{  
  Bubble-jets generating in tapered chips.(a) microchip on-line (top left) and sketch of internal jet formation within the gas-liquid two-phase flow in the tapered channel (middle). Region 1 is the main tapered zone,  and the inlet exhibits a rectangular cross-section ($h\times W_{\text{in}}$, where $h = 400\mu\text{m}$ and $W_{\text{in}} = W_{\text{out}} + 2X_L\tan(\alpha/2)$). The length of the tapered region is $X_L = 2600\mu\text{m}$. The outer diameter of the needle is $D = 400\mu\text{m}$, and the inner diameter is $d = 200\mu\text{m}$. The convergence angles are $\alpha = 5^{\circ}, 11^{\circ}, 17^{\circ}, 29^{\circ}$, and the needle displacements are $x = 500\mu\text{m}, 1000\mu\text{m}$, and $1500\mu\text{m}$. Region 2 is the parallel pipe flow  with a cross-section of $h\times W_{\text{out}} = 400\times 400\mu\text{m}^{2}$.(b) A typical image of sub-millimeter bubble jetting from 3D Volume of Fluid (VoF) simulations.(c) Experimental image corresponding to (b).
  }
   
  \label{fig:Schematic Diagram of the Convergent Microchannel}
\end{figure}

This letter reports our inventions of generating digital sub-millimeter bubble-jet, as well as relevant findings and laws. We designed polymethylmethacrylate (PMMA) micro-chips with tapered channels \cite{wang2015speed,wang2022universal}, as shown in Fig.\ref{fig:Schematic Diagram of the Convergent Microchannel}(a). The  unstable gas-liquid co-axial flow forms various flow patterns such as slug, jetting, dripping, and annular. The existence of the tapered zone breaks down the Galilean invariance to give two geometry variables: the taper angle $\alpha$ and the relative inner needle displacement $x$. In this work, $\alpha = 5^\circ, 11^\circ, 17^\circ, 29^\circ$, and $x = 500 \mu\text{m}, 1000 \mu\text{m}$, and $1500 \mu\text{m}$, respectively. Their combinations yield 12 experimental data sets.The gas (argon) is injected as the dispersed phase from the middle needle, while the continuous phase (deionized water) is injected from the outer channel. The flow rates of the continuous phase (deionized water) and the dispersed phase (argon) provided by a peristaltic pump and a mass flow controller ranges of 5 $\sim$ 100 mL/min and 5 $\sim$ 50 mL/min. The interfacial tension is 72.8 mN/m, and the density and viscosity of water are 986.2 kg/m³ and 1.23 mPa·s (measured at room temperature).The dynamics of the bubbles and jets are illuminated by a high-intensity LED light, and snapshots are captured at 20,000 Hz (Phantom V611-16G-M). The dynamic process of jet excitation within monodisperse bubbles can be clearly captured, as shown in Fig.\ref{fig:Schematic Diagram of the Convergent Microchannel}(c). We also used the Volume of Fluid (VoF) to compare and verify the corresponding experimental results (Fig.\ref{fig:Schematic Diagram of the Convergent Microchannel}(b)). 

Wang demonstrated that the co-axial liquid-liquid flow in the tapered region shows self-similarity and self-scaling. This was accomplished by introducing a similarity variable like the local virtual width, \cite{wang2022universal}.
\begin{equation} 
W_{\text{local}}=\frac{[2x+(2X_L tan(\alpha/2)+W_{\text{out}})cos(\alpha/2)]}{[1+sin(\alpha/2)]}-2x.
\label{eq:W_local}
\end{equation}
His data sets with different tapered configurations can be dimension-reduced to obtain a unified phase diagram and temporal and spatial scaling laws beyond physical mechanisms. We find that these self - similarity characteristics also hold in our gas-liquid system here.According to the corresponding conservation laws, the physical quantities are specially defined as follows: The superfacial velocity of the dispersed phase can be expressed as $u_d = Q_d / (\pi d^2 / 4) \cdot (W_{\text{local}}/W_{\text{out}})^{-1}$, while for the continuous phase is $u_c = Q_c / (W_{\text{local}} \cdot h - Q_d / u_d)$. The overall velocity is defined as $u_{\text{TP}} = (Q_{d}+Q_{c})/(W_{\text{local}}\cdot h))$. The corresponding capillary number of the continuous phase, Weber number of the dispersed phase, and overall capillary number of the two phases are defined as 
\begin{equation} 
Ca_c = \frac{\mu_c u_c}{\sigma}, We_d = \frac{\rho_d u_d^2 d}{\sigma}, and \; Ca_{\text{TP}} = \frac{\mu_c u_{\text{TP}}}{\sigma}.
\label{eq: dimensionless number}
\end{equation}
Under such definitions, since flow rate $Q_c$ and $Q_d$ ranges between 5 $\sim$ 100 mL/min and 5 $\sim$ 50 mL/min respectively, $u_d$ ranges 0.7 $\sim$ 4 m/s, and $u_c$ ranges 0.7 $\sim$ 7 m/s. Moreover, the range of $We_d$ is 0.01 $\sim$ 2, and that of $Ca_c$ is 0.01 $\sim$ 0.1.

\begin{figure}[!t]
  \centering
  \includegraphics[width=8.5cm,keepaspectratio]{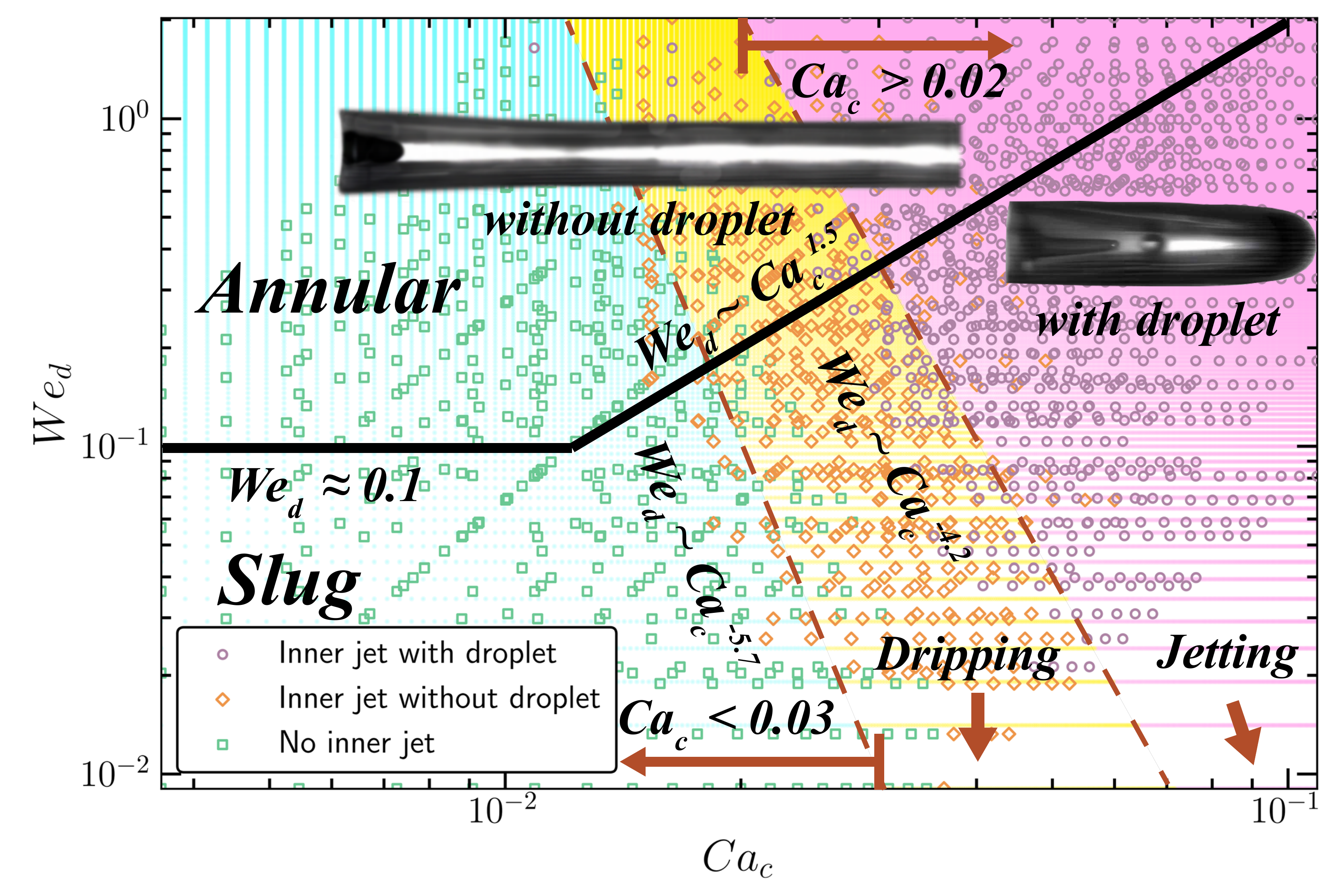}
  \caption{
  Phase diagram: flow pattern map orthogonally overlaping bubble-jets emitting chart. The thick black lines , where $We_d \approx 0.1$ for $Ca_c < 0.012$ and $We_d \sim Ca_c^{1.5}$ for $Ca_c > 0.012$, demarcate slug and annular. Dripping and jetting are absent in our scope (for $We_d \sim [0.01,2]$ and $Ca_c \sim [0.004,0.1]$). The two dashed lines, $We_d \sim Ca_c^{-5.7}$ and $We_d \sim Ca_c^{-4.2}$ give bubble-jetting line and jet-dropping line: no inner jet (Green, square), inner jet without droplets (Orange, diamond), and inner jet with droplets (Purple, circle).
  }
  \label{fig: Phase Diagram}
\end{figure}

After translating the data into self-scaling descriptions, we obtain a universal phase diagram in the $Ca_c \sim We_d$ space for all 12 different tapered configurations, as shown in Fig.\ref{fig: Phase Diagram}.Two types of phase charts are superimposed in Fig.\ref{fig: Phase Diagram}. One refers to flow patterns, and the other is the bubble-jet stimulating. Interestingly, these charts show obvious orthogonality and do not interfere, which indicating of  independent mechanism.Since the large flow-rate pumps are used and dimensioned by $ml/min$ rather than $ml/h$, the Reynolds number moderately ranges $[20 \sim 120]$ (Reynolds number is usually less than 1 in microfluidic research). Therefore, only two flow patterns, slug and annular, appear within this parameter range. The slug, known as Taylor flow, is regular periodic bubble flows, which relatively easy to identify (at the bottom of the phase diagram). While the annular is actually two types. One is the case where the gas-liquid flow is always stratified without disperse phase collapse (upper-left), where the inertia of the dispersed phase is relatively large while the viscosity of the continuous phase is relatively small. The other type of annular flow is formed during bubbles chase and coalescence, constantly merging and splitting into dynamic long bubbles (upper-right), where the inertia of the dispersed phase is very large, and the viscosity of the continuous phase is also relatively large. The dispersed phase will collapse. 

\begin{figure}[!t]
  \centering
  \includegraphics[width=8.5cm,keepaspectratio]{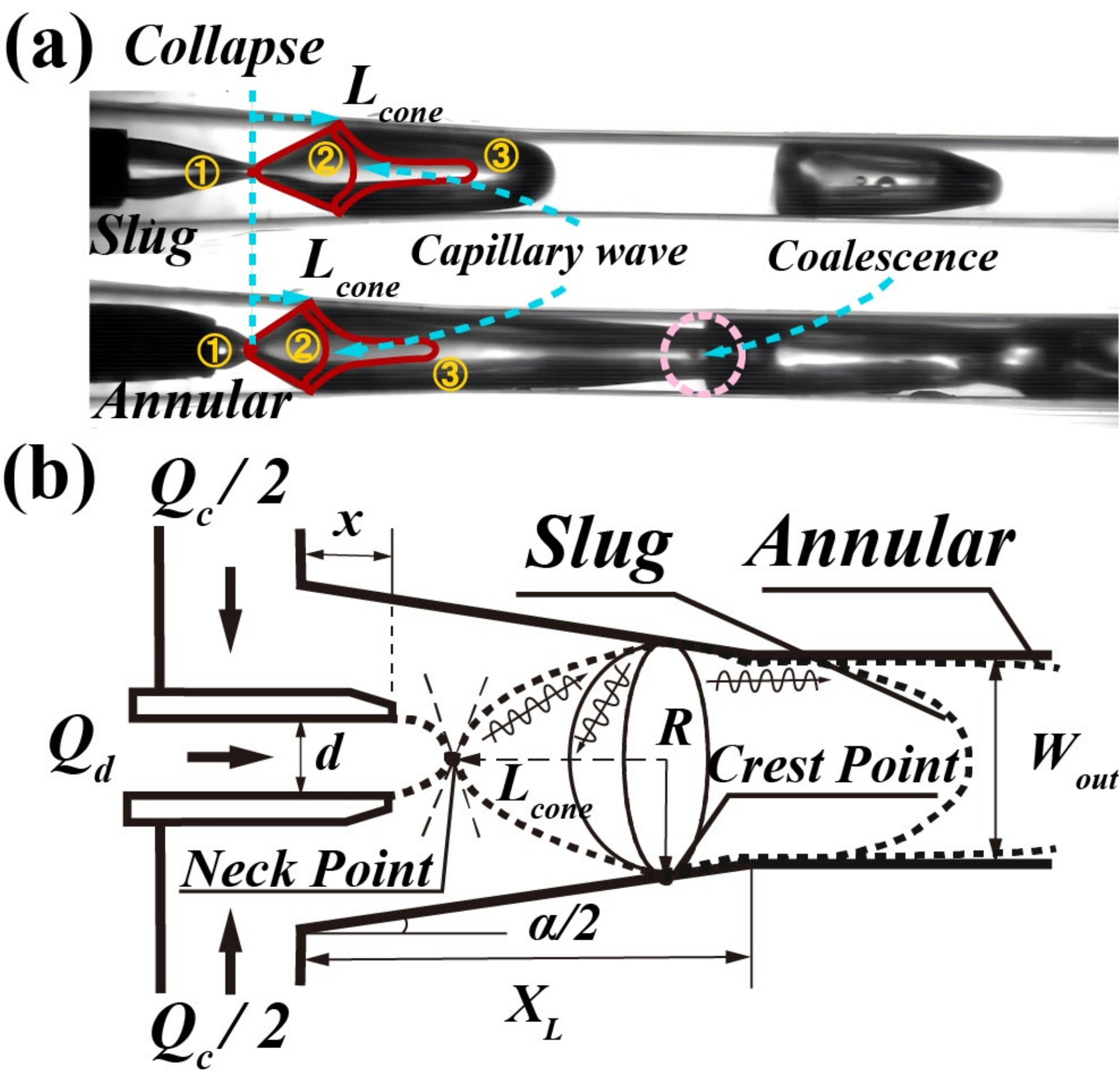}
  \caption{
  Onsets of Ejections: a local model.(a) The jet formation processes of both slug (upper) and annular (lower) involve three stages (indicated by the red contour): (1) interface collapse, (2) propagation and focusing of interfacial capillary waves, and (3) jet formation. (b) Schematic of the local model for onsets of bubble-jetting. We summarize the slug and annular jet formation processes as local flow behaviors within the tapered region. The transmission of capillary waves (wavy arrows in the figure) at the interface predominantly occurs at the bubble's tail. Consequently, a local measured quantity, $L_{\text{cone}}$, is introduced as an intermediate variable for analysis. $L_{\text{cone}}$ is defined as the axial distance from the neck point to the bubble's maximum circumferential diameter (Cone Crest).
  }
  \label{fig: Motion Model Diagram}
\end{figure}

And bubble-jets only occur when the dispersed-phase gas collapse.In fact, we found that there are three situations for bubble-jets to occur: 1) Dispersed-phase collapse near the nozzle in the tapered region; 2) Generated by the chasing and merging of two bubbles (coalescence in the annular snapshot in Fig.\ref{fig: Motion Model Diagram}(a)); 3) Inward-rolling of a long bubble during its movement \cite{pico2024drop}. This topic relates the first case.According to Fig.\ref{fig: Phase Diagram}, two types of threshold values exits for bubble-jets: 1) The bubble-jets excitation threshold; 2) The jet droplet emitting threshold. To generate an internal jet, a relatively large momentum transfer is required, and the shear force of the continuous phase needs to be large. However, the role of surface tension is rather contradictory. On the one hand, it suppresses surface deformation, on the other hand, it provides the original driving force for bubble-jet emitting.Actually, The threshold for the bubble-jets emitting has been discussed extensively in problems like free-surface emergence of bubble-jet, which depend on the critical Ohnesorge number ($Oh=\mu / \sqrt{\rho R \sigma}$, where $R$ is the initial bubble radius) and is related to the Bond number \cite{ganan2017revision}. As can be seen from our phase diagram Fig.\ref{fig: Phase Diagram}, this critical value also exists, manifested as the oblique dotted line $We_d \sim Ca_c^{-5.7}$ (with a lower intercept of $Ca_c \approx 0.03$). Similarly, the threshold value for the bubble-jet to generate droplets also exists, manifested as $We_d \sim Ca_c^{-4.2}$ (with an upper intercept of $Ca_c \approx 0.03$). We can define dimensionless numbers again to give the corresponding critical constants, but this will lead to problems such as inconsistent definitions and unclear physical concepts.Obviously, these existing thresholds are caused by the competitive relationships among various physical characteristics. Here, the dominant factor is the comparison among shear force, inertial force, and surface tension. However, this relationship can dominate the characteristics of the flow-pattern phase diagram. How does it transfer to the generation mechanism of bubble-jets? Obviously, the process of the bubble-jets generating droplets follows the well-studied Rayleigh-Plateau jet stabilities \cite{yu2023bubble, anna2016droplets, castillo2015droplet, lhuissier2013drop}. Therefore, the remaining most crucial issue is to clarify the initial bubble-jet excitation mechanisms and the laws of the intensity of bubble-jets.

From an observational perspective, the choices of what we can measure are of great significance. Regarding the excitation intensity of bubble-jet, following common practice, the tip velocity $u_{\text{jet}}$ can be selected to represent \cite{seon2012large,ghabache2014physics,ganan2017revision,ganan2018scaling, deike2018dynamics,li2020jet}. Since the period for bubble-jet emitting is extremely short, less than 0.2 milliseconds, which makes the tip velocity change significantly and easy to capture, and suitable for observing morphological evolution processes. We measured $u_{\text{jet}}$ ranges $1 \sim 13 m/s$. Noted that, different from the jet triggered at the free surface, our bubble is moving, $u_{\text{jet}}$ we refer to is relative to the overall movement of the bubble. Thus, the actual jet velocity ranges approximately $3 \sim 25 m/s$, which the relative velocity $u_{\text{jet}}$  can better reflect the characteristics of the jet's excitation intensity. Therefore, macroscopically, we can write the expression, 
\begin{equation} 
u_{\text{jet}}=F(\rho_d, \rho_c, \mu_d, \mu_c, \sigma; h, W_{\text{out}}, D, d, W_{\text{local}}; Q_d, Q_c). 
\label{eq:u_jet-F}
\end{equation}
The dependent variables of $u_{\text{jet}}$ fall into three categories: physical property parameters, geometric parameters, and flow control parameters.This expression makes no difference from cavernous descriptions of digital bubble generating \cite{wang2022universal}. Though representing the characteristics of the tapered region with the similarity variable \(W_{\text{local}}\) has achieved significant dimensionality reduction, there are still quite a number of variables. As a result, the experimental cost is too high to withstand. Of course, many parameters can be regarded as constants, but doing so will lose the generalization of the research.

Let's first analyze the bubble-jet emitting process, as shown in Fig.\ref{fig: Motion Model Diagram}(a). Through observation, we can find the following: 1) The break-up of the dispersed-phase occurs in the tapered region, where presents a “front-end stretching and rear-end squeezing” accelerated stretching flow \cite{wang2015speed}. 2) The bubble-jet emitting of slug and annular are similar or identical. Thus, macroscopic bubble characteristics do not influence the mechanism and intensity of bubble-jet emitting. This explains why there is orthogonality between the flow-pattern map and the jet-emitting chart in Fig.\ref{fig: Phase Diagram}. 3) During the processes of collapse, capillary-wave propagation, focusing, and bubble-jet formation (the red line in Fig.\ref{fig: Motion Model Diagram}(a)), the interface vibration only takes place in the tail region of the bubble. These observations suggest that the action, formation, and triggering of bubble-jet should be a local process.If we develop a local model, it should be reduced to Fig.\ref{fig: Motion Model Diagram}(b). We only need to consider the region from the jet break-up point (Neck point) to the point of the bubble's maximum diameter (Crest point), while the influence of the bubble's head (the parts of slug and annular drawn with dotted lines) can be ignored. Here, we've identified another measurable quantity, \(L_{\text{cone}}\), as seen in Fig.\ref{fig: Motion Model Diagram}(b). Through this intermediate variable \(L_{\text{cone}}\), we may bridge the overall macroscopic bubble behavior with the local microscope of jet emitting. 

\begin{figure}[!t]
  \centering  \includegraphics[width=8.5cm,keepaspectratio]{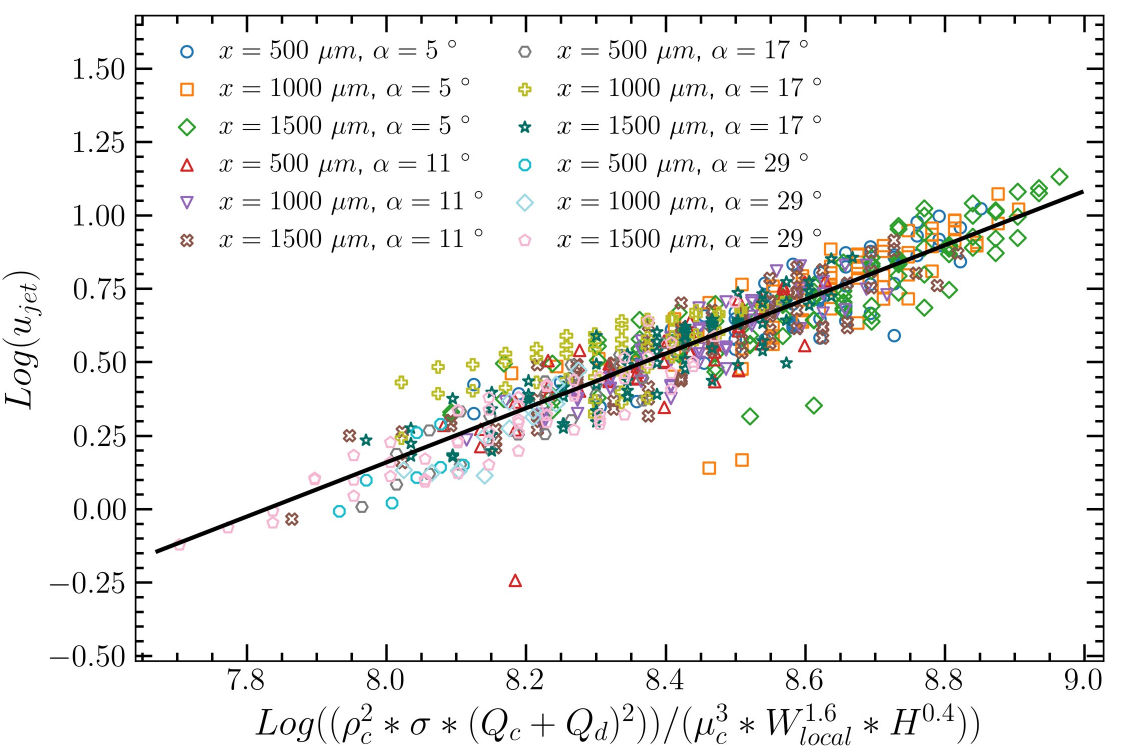}
  \caption{  
 $\log(u_{\text{jet}}) \sim \log\left(\frac{\rho_c^2 \sigma (Q_c + Q_d)^2}{\mu_c^3 W_{\text{local}}^{1.6} H^{0.4}}\right)$. Symbols represent data sets from all 12 datasets ($x = 500 \mu\text{m}, 1000\mu\text{m}, 1500 \mu\text{m}$ combined with $\alpha = 5^\circ, 11^\circ, 17^\circ, 29^\circ$) where bubble-jets fully occur.  
  }
  \label{fig: The relationship between jet velocity and initial variables}
\end{figure}

$L_{\text{cone}}$ represents the extreme stretching state of the bubble tail and serves as the initial configuration for onset of ejections, thus it holds great significance. To find mechanisms in bubble tail stretching, we will look into the relevant energy processes. Taking into account the previous analysis, the substantial stretching of the bubble is associated with the structure of the tapered region for the “front-end stretching and rear-end squeezing” effects  \cite{wang2015speed}. Unlike the situation of a freely breaking jet, the stretching in this non-uniform shear flow field is much more pronounced. The work $W$ done by the fluid on the gas-liquid interface can be expressed as, 
\begin{equation}
W \approx W_{sq}+W_{st} =W_{sq1}+W_{sq2}+W_{st}.
\label{eq:power}     
\end{equation} 
Among them, the work $W_{sq}$ done on the gas-liquid interface by “rear-end squeezing” should come from the work done in the direction of interface deformation by the normal pressure difference $\Delta p \sim \sigma \kappa+\rho_c a W_{\text{local}}$ across the interface and the shear stress $\tau \sim \mu_c u/W_{\text{local}}$ along the interface tangent. They can be expressed as $W_{sq1}=\int_S \Delta p(\mathbf{u} \cdot \mathbf{n}) \mathrm{d} A$ and $W_{sq2}=\int_S \tau(\mathbf{u} \cdot \mathbf{n}) \mathrm{d} A$ respectively. Here, $\kappa \sim W_{\text{local}}$ is the interface curvature, $u \sim (Q_d + Q_c)/W_{\text{local}}^2$ is the flow field velocity, $a = du/dt \sim u^2/W_{\text{local}}$ is the fluid acceleration caused by tapering, and $S \sim W_{\text{local}}^2$ is the area of the bubble-tail interface on which work is done.The work $W_{st}$ on the interface of the bubble head described by the “front-end stretching” effect is essentially the same as $W_{sq}$, so the work expressions are similar, with the difference lying in the specific flow-field characteristics. However, since the bubble-jet we observed exhibits the characteristics of local flow behavior, and the occurrence time of the entire bubble-jet process is much shorter than the flow time, the work of the overall “front-end stretching” of the bubble should be negligible in the triggering process of the bubble-jet, unless the bubble-jet occurs in the case of relatively small droplets (not belonging to the slug and annular flow patterns). Then we have $W_{st}\approx 0$, and thus $W \approx W_{sq}=f(Q_d + Q_c,W_{\text{local}};\rho_c,\mu_c,\sigma)$.As an observable quantity representing the result of work-done, the expression of the intermediate variable $L_{\text{cone}}$ should also contain the same parameter variables. Therefore, we write $L_{\text{cone}}$ as the following formula, 
\begin{equation} 
L_{\text{cone}}=F_{1}(Q_d+Q_c,W_{\text{local}};\rho_c,\mu_c,\sigma). 
\label{eq:L_cone-F1}
\end{equation}
Taking $W_{\text{local}}$, $\sigma$, and $Q_d + Q_c$ as independent characteristic variables, we obtain the dimensionless expression of Eq.\ref{eq:L_cone-F1}, that is $L_{\text{cone}}/W_{\text{local}} \sim  \Phi (Oh_{\text{local}},Ca_{\text{local}})$. Among them, $Oh_{\text{local}} = \mu_c / (\rho_{c} \sigma W_{\text{local}})^{0.5}$ and $Ca_{\text{local}} = \mu_c (Q_d + Q_c) / (\sigma \cdot W_{\text{local}}^2)$. These two dimensionless numbers jointly characterize the local behavior of bubble stretching. $Oh_{\text{local}}$ emphasizes the competitive relationship between viscous force, inertial force, and surface tension, while $Ca_{\text{local}}$ describes the competitive relationship between fluid inertia and surface tension. They play a crucial role in the morphological deformation of the bubble, the jet velocity, and the final jet intensity. Through experimental data analysis, we obtain the relationship: $L_{\text{cone}}/ W_{\text{local}} \sim Oh_{\text{local}}^{-2} \cdot Ca_{\text{local}}^{1.6}$. Restoring it to the form of original variables, we get 
\begin{equation}
L_{\text{cone}}\sim \frac{\rho_c (Q_d + Q_c)^{1.6}}{\sigma^{0.6}\mu_c^{0.4}W_{\text{local}}^{1.2}}
\label{eq:L_cone-Q}
\end{equation}
Formula \ref{eq:L_cone-Q} is in excellent agreement with our experimental data. For details, you can refer to the supplementary materials. Now that we've clearly analyzed the initial morphological characteristics in the onset of ejections process, let's move on to discuss the intensity and mechanism of bubble-jet. Since the bubble-jet emitting process is a transformation from potential energy to kinetic energy, this transformation mainly involves two parts. One is the restoring potential energy of the stretched bubble, and the other is the energy obtained by the jet from the change in inertia of the two-phase fluid. Therefore, we have 
\begin{equation}
K_{in} \approx \delta E_{ca}+\delta E_{in}-E_{dis}. 
\label{eq:power}     
\end{equation} 
Among them, $K_{\text{in}}=\int_V 1/2 \rho_c u_{\text{jet}}^2 dV$ represents the kinetic energy of the jet, where $u_{\text{jet}}$ is the field velocity which is impractical for experiments. In fact, we usually substitute the field velocity in the integral with measured tip velocity of the jet, thus $K_{in} \sim \rho_c u_{\text{jet}}^2$. Since we are discussing the initial energy balance of the bubble-jet here, the intermediate variable $L_{\text{cone}}$ is an excellent local spatial characteristic quantity to reflect the energy situation during the bubble-jet triggering stage. Therefore, the capillary restoration energy is $\delta E_{ca} \sim \sigma L_{\text{cone}}^2$, and the inertial energy change is $\delta E_{in} \sim \rho_c u_{\text{TP}}^2$. Because the time of bubble rebound is shorter compared to the acceleration effect in the tapered region of the flow field, the viscous dissipation $E_{dis} \sim \mu_c (u_{\text{TP}}/L_{\text{cone}})^2$ should be taken into account during the rebound process and the bubble-jet emitting process. In this way, from Eq.\ref{eq:power}, the expression of $u_{\text{jet}}$ can be written as follows,
\begin{equation} 
u_{\text{jet}}=F_{2}(L_{\text{cone}},u_{\text{TP}};\rho_c,\mu_c,\sigma).
\label{eq:u_jet-F2}
\end{equation}
Select $\rho_c$, $\sigma$, and $u_{\text{TP}}$ as independent scaling variables, we derive $u_{\text{jet}}/u_{\text{TP}}\sim f(We_{\text{TP}},Ca_{\text{TP}})$. Among them, $We_{\text{TP}}= \rho_c L_{\text{cone}} u_{\text{TP}}^2/ \sigma$ represents the relative importance between the inertial force and the surface tension determined by $u_{\text{jet}}$ and $L_{\text{cone}}$ during the local similarity. Through experimental data analysis, we obtain the relationship $u_{\text{jet}}/u_{\text{TP}} \sim We_{\text{TP}} \cdot Ca_{\text{TP}}^{-2.6}$. Finally, when expanded into the primitive variables, we have 
\begin{equation} 
u_{\text{jet}}\sim \frac{\rho_c \sigma^{1.6}u_{\text{TP}}^{0.4}L_{\text{cone}}}{\mu_c^{2.6}}. 
\label{eq:u_jet-u}
\end{equation}
It is evident that the intermediate variable $L_{\text{cone}}$ exhibits a remarkable connective effect. Upon substituting Eq. \ref{eq:L_cone-Q} into Eq. \ref{eq:u_jet-u}, the final relationship of the bubble-jet velocity $u_{\text{jet}}$ can be expressed as,
\begin{equation} 
u_{\text{jet}}\sim \frac{\rho_c^2 \sigma (Q_c + Q_d)^2}{\mu_c^3 W_{\text{local}}^{1.6} H^{0.4}}.
\label{eq:u_jet-Q} 
\end{equation}
Thus, the global expression of $u_{\text{jet}}$ in Eq.\ref{eq:u_jet-F} is solved, and answers the jet intensity problem. Plot Eq.\ref{eq:u_jet-Q} in Fig.\ref{fig: The relationship between jet velocity and initial variables}, it can be seen that the 12 datasets are basically on the formula (black straight line),which is quite accurate in characterizing the bubble-jet intensity features.

Eq.\ref{eq:u_jet-Q} is fascinating for: 1) Incorporating inertial, surface tension, viscous, and the tapered geometry effects, which is consistent with viewpoints that the magnitudes of various mechanical quantities at the bubble-jet triggering stage are similar \cite{ganan2017revision, ismail2018controlled}. 2) Comparing $W_{\text{local}}^{1.6}$ and $H^{0.4}$, the effect of the tapered scale is to the fourth power of the pipe width, indicating that the fluid acceleration has significant influences on jet emitting, agreeing with literature \cite{wang2015speed, wang2022universal}. 3) If transformed, we can have $u_{\text{jet}}\propto Re^3/La$, where $La=\frac{\rho U^2 L}{\mu \sigma}$, which shows fluid inertia is crucial in bubble-jet occurrence; also for bubble-jet digital emitting not being discovered in microfluidic researches due to small $Re$. 4) Surface tension is still positive, which is basically consistent with works on bubble emergence, bubble chasing, etc. \cite{brasz2018minimum, berny2020role, andredaki2021accelerating}. 5) Note that the formula holds for all the tapered configurations, and its generality can be ensured, that is, the tapered gas-liquid flow maintains self-scaling flow structures. 6) Such regular exponents from experimental data under compound physical mechanisms is rare, which indicates that the competing among various mechanisms is evenly matched, acting clearly, so the data can be well-posed for physical laws. 7) Due to the limits of instruments in ranges ($We_d \sim [0.01, 2]$, $Ca_c \sim [0.004, 0.1]$), outer border of the bubble-jetting is open, and critical boundary is unclear, when bubble-jet velocity of magnitude of $O(10^2)$ appears in our simulations, which leads to another interesting topic.

Another thing that needs to be clarified is the locality issue. For the local model in Fig.\ref{fig: Motion Model Diagram}(b), it draws the forward capillary wave propagating. So how is the local nature of the model ensured? Three clues: 1) According to free space bubble-jetting, where capillary waves move towards the bubble bottom and focusing, called inertial-capillary focusing \cite{ghabache2014physics,andredaki2021accelerating,berny2020role}, our waves should meet at the bubble head and excite jets. But here the bubble-jetting occurs at the bubble tail, so the jet must be triggered by reflected waves. When the capillary wave reaches crest points, it encounters the boundary resistance and quickly reflects, then focuses at the channel center and triggers the jet. Therefore, the reflected wave is dominant. 2) Capillary wave speed $c=\sqrt{(\sigma / \rho) k}$($k$ wave wavelength $\sim 1mm$), $\sim 0.3m/s$. The time for wave to reflect and focus after reaching the crest point is $W_{local}/(2c)$ or $h/(2c)$, $\sim 1 \mu s$. For the forward wave, this time is insufficient to travel out of tail region. 3) In terms of intensity, the capillary wave may be relatively weak to give limited energy. But, for exciting a fast jet, the role of the inertial wave is large, which is shown in Eq.\ref{eq:power} and  Eq.\ref{eq:u_jet-Q}. Usually, the inertial wave speed is much greater than that of the capillary wave, which shorts the triggering time, also the chance of forward wave leaving.

This work was supported by the National Natural Science Foundation of China - Yunnan Joint Fund Key Project, U2002214. The authors would like to express their sincere gratitude to Professors Jin-Song Zhang, Guo-Hui Hu, Zhe-Wei Zhou, and Qian Xu for valuable assistance.

\bibliography{apssamp}

\end{document}